\documentclass[twocolumn,aps,preprintnumbers,amsmath,amssymb]{revtex4}
\pdfoutput=1
\usepackage{graphicx} 
\usepackage{amsmath}
\usepackage{amssymb}
\usepackage{mathrsfs}
\usepackage{subfig}
\usepackage{graphicx}
\usepackage{dcolumn}
\usepackage{bm}
\usepackage{color}
\begin{document}

\title{Confronting  the Stochastic Neutrino Mixing Mechanism and the sterile neutrino hypothesis as a solution to the short baseline neutrino anomalies}

\author{E. M. Zavanin\footnote{Corresponding author: zavanin@ifi.unicamp.br}}
\affiliation{Instituto de F\'isica Gleb Wataghin - UNICAMP, {13083-859}, Campinas SP, Brazil}

\author{M. M. Guzzo}
\affiliation{Instituto de F\'isica Gleb Wataghin - UNICAMP, {13083-859}, Campinas SP, Brazil}

\author{P. C. de Holanda}
\affiliation{Instituto de F\'isica Gleb Wataghin - UNICAMP, {13083-859}, Campinas SP, Brazil}

\author{O. L. G. Peres}
\affiliation{Instituto de F\'isica Gleb Wataghin - UNICAMP, {13083-859}, Campinas SP, Brazil}
\affiliation{
The Abdus Salam International Centre for Theoretical Physics, I-34100 Trieste,
Italy
}

\date{\today}

\begin{abstract}

We compare the solutions to the short baseline neutrino anomaly based on oscillations to sterile neutrinos and the Stochastic Neutrino Mixing Mechanism (SNMM) through an analysis of the present neutrino data. The SNMM suggests worse fits than a 3 + 1 sterile neutrino model, although it cannot be discarded by present data. We propose an experiment to distinguish between both solutions, based on placing a  $^8$Li source inside a 5kton-yr detector (like SNO). We studied the sensitivity of such an experiment, which makes it possible to discriminate within $2\sigma$ the SNMM from the 3+1 sterile  hypothesis  for some particular values of the relevant parameters  in 5 kton-years of running.

\end{abstract}

\maketitle

\section{Introduction}

The recent analysis provided by Ref.~\cite{PhysRevC.83.054615} increased the theoretical prediction of the reactor anti-neutrino mean flux from $^{235}$U, $^{239}$Pu, $^{241}$Pu, and $^{238}$U by about 3 percent. The published reactor experiments at short baseline ($<$ 100 m) reported a ratio of observed event rate over predicted rate of 0.976~$\pm$ 0.024. In the new approach, this ratio shifts to 0.943~$\pm$ 0.023, leading to a deviation from unity at 2.5$\sigma$. This is called the reactor anti neutrino anomaly~\cite{PhysRevD.83.073006}.  The ratio of observed over predicted events was also studied with an intense artificial $^{51}$Cr and $^{37}$Ar radioactive sources, which were placed inside the detectors of GALLEX~\cite{Anselmann1995440,Hampel1998114} and SAGE~\cite{PhysRevC.80.015807}, and also indicated rates smaller than unity of $0.86 \pm 0.05$, which has a deviation from the unity at 2.8~$\sigma$. This is called the Gallium anomaly~\cite{PhysRevC.83.065504}. 

Moreover, searches for $\bar{\nu}_{\mu}\rightarrow \bar{\nu}_e$ oscillations conducted by the Liquid Scintillator Neutrino Detector (LSND) found an excess of events above the expected background~\cite{PhysRevD.64.112007}. The MiniBooNE experiment also reported an excess of events at 3.8~$\sigma$, combining the $\nu_e$ and $\bar{\nu}_e$ data sets~\cite{Aguilar-Arevalo:2013pmq}.

This panorama cannot be explained by the standard model of neutrino oscillations, in which only active neutrinos are allowed to oscillate. Once the oscillation length related to the mass eigenstates 1-2 is well measured by Solar and KamLand~\cite{PhysRevLett.94.081801} data and the oscillation length of  mass eigenstates 1-3 is also well established by atmospheric neutrinos~\cite{Ashie:2004mr} and long baseline experiments~\cite{Adamson:2014vgd,Abe:2014ugx}, the excess in MiniBooNE, LSND, old reactors and Gallium anomalies, cannot be accommodated in a three oscillating neutrino families context.

To explain such anomalies, the neutrino sterile hypothesis is the most investigated scenario~\cite{Abazajian:2012ys}. Sterile neutrinos are additional states beyond the standard electron, muon, and tau flavors, which do not interact by charged currents (CC) neither by neutral currents (NC). Sterile neutrinos are connected to additional mass states via an extended mixing matrix with extra mixing angles.

Other proposals to handle such anomalies were also studied by several authors, like CPT violation~\cite{1742-6596-335-1-012054} and extra dimensions~\cite{PhysRevD.85.073012}. In special, a proposal was made in~\cite{snmm}, based on what we called Stochastic Neutrino Mixing Mechanism (SNMM) in which the mixing angles are stochastic variables that can  be different in the process of neutrino creation and detection. We argued that such mechanism seems to supply a possible explanation to all these anomalies. 

In this paper we will improve the analysis made in~\cite{snmm} adding the MiniBooNE data and making a spectrum analysis of the accelerator data. We also will compare the SNMM with the sterile hypothesis through a statistical analysis.

We start presenting the 3+1 sterile neutrinos model analysis~\cite{Peres:2000ic}. Then we introduce the foundations of the SNMM and the dependence of their free parameters with the energy. We compare, through a statistical analysis, the fits of sterile neutrinos and the SNMM. In the last section we study a $^8$Li source as a possible experimental test to distinguish the solutions to the short baseline anomalies based on the SNMM and sterile neutrino hypothesis.

\section{Framework of Sterile Neutrino and Stochastic Neutrino Mixing Mechanism}

\subsection{Sterile Neutrinos}

While the indications of sterile neutrino oscillations have historically been associated only with appearance-based short baseline experiments~\cite{PhysRevD.64.112007}, the recently observations of disappearance of $\nu_e$ and $\bar{\nu}_e$~\cite{PhysRevD.64.112007,PhysRevC.83.065504}, and new data reporting appearance of $\bar{\nu}_e$~\cite{Aguilar-Arevalo:2013pmq}, provides further motivation for these models. The more economical model is the so-called 3+1 model~\cite{Peres:2000ic}, which involves a fourth neutrino states, sterile, which relates to the usual active eigenstates by a $4\times 4$  mixing matrix, $\nu_{\alpha}=U_{\alpha i} \nu_i$ where i=1,2,3,4. 

Assuming very-short baselines, where the only relevant mass-scale is the one involving the fourth neutrino family, the flavor oscillation probabilities in sterile neutrino case are given by:
\begin{eqnarray}
P (\nu_e \rightarrow \nu_e) = 1 - U_{e 4}^4 \sin^2(2\theta_{13}) \sin^2\left(\frac{\Delta m^2_{31} L}{4E}\right) \\ \nonumber 
- 4 U_{e 4}^2(1 - U_{e 4}^2)\sin^2\left(\frac{\Delta m^2_{41} L}{4E}\right),
\label{steril1}
\end{eqnarray}
for the electronic neutrino survival probability, and: 
\begin{equation}
P_{\nu_\mu \rightarrow \nu_e} = 4 U_{e 4}^2 U_{\mu 4}^2\sin^2\left(\frac{\Delta m^2_{41} L}{4E}\right).
\label{steril2}
\end{equation}
for the transition probability.  Here $L$ is the neutrino traveled distance, $E$ is the neutrino energy. Therefore, all oscillation data of short baseline experiments can be described by adding these parameters: $U_{e4}^2$  and  $U_{\mu4}^2$ elements of the mixing matrix and  the mass scale $\Delta m^2_{41}$.

Nevertheless, introducing sterile neutrinos can have implications in cosmological observations, especially in measurements of the radiation density in the early universe if the extra neutrinos have significant mass ($>$1 eV) and do not decay. Recently, Ref.~\cite{Ade:2013zuv} estimates the effective number of neutrinos to be $N_{\rm eff} = 3.30 \pm 0.27$, that indicates disfavored limits to sterile neutrinos. A complete analysis of global fits to sterile neutrinos was obtained in~\cite{Conrad:2012qt}. Despite the fact that these fits provide excellent results to the anomalies, it is evident that the great values associated with these sterile states mass are in conflict with cosmological bounds~\cite{Hamann:2011ge}. Having in mind this incompatibility of the sterile neutrino scenario with cosmological data, besides the fact that there is no direct evidence of sterile neutrinos, we start studying another proposal which does not add any new massive state, therefore not violating cosmological constraints: the SNMM.

\subsection{Stochastic Neutrino Mixing Mechanism}

The Stochastic Neutrino Mixing Mechanism make two assumptions:
\begin{enumerate}
\item the mixing angles that compose the neutrino eigenstates at creation and detection can be different;
\item the mixing angles are stochastic variables.
\end{enumerate}
The first assumption can generate a non-zero oscillation probability even for zero distance,  which is the central point to explain the data related to short baseline anomalies. This is similar to non-unitary oscillation mechanism described in Ref.~\cite{Ohlsson:2012kf}.

 The  stochastic nature of our mechanism appears as we integrate over the stochastic variables $\theta_c$ and $\theta_d$, which are the neutrino mixing angle at the creation and detection respectively. The total oscillation probability becomes: 
\begin{equation}
<P (\nu_\alpha \rightarrow \nu_\beta)> = \int_0^{\pi/2}\int_0^{\pi/2} P_{\nu_\alpha \rightarrow \nu_\beta}f(\theta_c)f(\theta_d)d\theta_c d\theta_d,
\label{Preal3fam1}
\end{equation}
where the $<>$ symbols represent the averaging due to the stochastic mechanism.
Here $\alpha$ and $\beta$ denotes the neutrino flavor $e$, $\mu$ or $\tau$, $f(\theta)$ is a distribution function to be defined below, and:
\begin{eqnarray}
P(\nu_\alpha\rightarrow \nu_\beta)= \sum_{\gamma}{|U^{c}_{\alpha\gamma}|^2 |U^{d}_{\beta\gamma}}|^2 \\ \nonumber
+ 2\sum_{\gamma>\delta}{U^{c}_{\alpha\gamma}U^{d}_{\beta\gamma}U^{c}_{\alpha\delta}U^{d}_{\beta\delta} \cos\left(\frac{\Delta m^2_{\gamma\delta} L}{2E}\right)},
\label{snmm3fam1}
\end{eqnarray}
where $\gamma$ and $\delta$ run from 1 to 3 and $\Delta m^2_{\gamma\delta}$ is the usual squared mass difference between the mass eigenstates involved in the oscillation process. 

In order to parametrize this different mixing angles in the creation and detection procedure we define a $3\times 3$ mixing matrix at the moment of the neutrino creation ($U^{\rm c}$) and at the detection moment ($U^{\rm d}$), in the following way: 
\begin{eqnarray}
\small
U^{{\rm c,d}} =\left(
 \begin{array}{ccc}
  c_{12}^{\rm c,d} c_{13} & s_{12}^{\rm c,d} c_{13} & s_{13} \\
  -s_{12}^{\rm c,d} c_{23} - c_{12}^{\rm c,d} s_{23} s_{13} & c_{12}^{\rm c,d} c_{23} - s_{12}^{\rm c,d} s_{23} s_{13} & s_{23} c_{13} \\
  s_{12}^{\rm c,d} s_{23} - c_{12}^{\rm c,d} c_{23} s_{13} & -c_{12}^{\rm c,d} s_{23} - s_{12}^{\rm c,d} c_{23} s_{13}  & c_{23} c_{13}  \\
   \end{array}\right)
\nonumber \nonumber
\end{eqnarray}
where $c_{ij}=\cos \theta_{ij}$, $s_{ij}=\sin\theta_{ij}$,  the stochastic variables are $c_{12}^{\rm c,d}=\cos\theta_{12}^{\rm c,d}$ and $s_{12}^{\rm c,d}=\sin\theta_{12}^{\rm c,d}$, and $\theta_{12}^{c,d}$ can assume values in the interval $[0,\pi/2]$. For the general case in which $\theta_{12}^c\neq\theta_{12}^d$, the survival probability, Eq.~(\ref{Preal3fam1})  is less than unity even at zero distance, also the conversion probability is non-zero for zero distance. Such behavior, which is not allowed in the usual oscillation processes, is the essence of the SNMM.  For simplicity, we choose the 
stochastic variables distribution function as a gaussian distribution:
\begin{equation}
f(\theta_{12}^{c,d}) = \frac{1}{\sqrt{N}}\exp{\left[-\left(\frac{\theta_{12}^{c,d}-\theta_{12}}{\alpha}\right)^2\right]},
\label{dist}
\end{equation}
where  the $N$ is the normalization factor.  We assume for $\alpha$ an energy dependence given by  $\alpha = A + B/E^n$, with $A$, $B$ and $n$ as free parameters. 

In order to clarify the model we can write explicitly the conversion 
probabilities
replacing the coefficients:
\begin{eqnarray}
&&v_{\alpha,\gamma\delta}=<U_{\alpha\gamma}U_{\alpha\delta}> \nonumber \\
&=&\int d\theta_{12}^cf(\theta_{12}^c)U_{\alpha\gamma}^cU_{\alpha\delta}^c=
\int d\theta_{12}^df(\theta_{12}^d)U_{\alpha\gamma}^dU_{\alpha\delta}^d
\label{averaging}
\end{eqnarray}
in Eqs.~(\ref{Preal3fam1}) and (\ref{snmm3fam1}), obtaining:
\begin{eqnarray}
 & & <P(\nu_\alpha\rightarrow\nu_\beta)>= 
\sum_\gamma v_{\alpha,\gamma\gamma}v_{\beta,\gamma\gamma} \nonumber \\
+& & 2\sum_{\gamma>\delta} v_{\alpha,\gamma\delta}v_{\beta,\gamma\delta}
\cos\left(\frac{\Delta m_{\gamma\beta}^2}{2E}L\right), 
\label{snmm3fam1IIf}
\end{eqnarray}
where we see that the stochastic effect works as a modulation in the probability even in very short-baselines distances. The second term in Eq.~(\ref{snmm3fam1IIf}) modulates the amplitude of oscillation, but not destroying the pattern of oscillation which is important to describe the accelerator, reactor and the atmospheric data.

In order to perceive the changes on flavor conversion induced by the stochastic mechanism, is particularly clear to analyze the electronic survival probability at very short baselines in the approximation $\theta_{13}=0$. Making $L\rightarrow 0$ in Eq.~(\ref{snmm3fam1IIf}), under this approximation, we can rewrite the survival probability as:
\begin{eqnarray} 
&<P(\nu_e \rightarrow\nu_e)> =\sum_\gamma v_{e,\gamma\gamma}^2
+2\sum_{\gamma>\delta} v_{e,\gamma\delta}^2\\ \nonumber
&=1+2\left[<c_{12}s_{12}>^2-<c_{12}^2><s_{12}^2>\right]
\end{eqnarray}
where the $<>$ symbols represent the averaging due to the stochastic mechanism, like in Eq.~(\ref{averaging}). 
From this expression is easy to see that if the stochastic mechanism is not present, due to unitarity the survival probability at very small baseline is $1$. So, the effect of stochastic neutrinos is a zero distance effect, allowing flavor conversions for very short baselines.

For the solar neutrinos, the averaging over oscillation takes place, and only the first term in 
Eq.~ (\ref{snmm3fam1IIf}) provides flavor conversion. And for this term we should analyze the stochastic mechanism effects in two different energy regimes. For high energy solar neutrinos, the matter effect dominates over the oscillation term $\Delta m^2/4E$, and all neutrinos are produced as mass eigenstate $\nu_2$. As a result, the averaging over the stochastic angle is not effective and the only effect of the stochastic mechanism would be an averaging over a mixing angle on detection point, leading to a negligible effect on probability. For low energy solar neutrinos, the matter effect are not large enough to change the mixing angles, and we obtain the first term of Eq.~(\ref{snmm3fam1IIf}) averaged over production and detection mixing angles. We numerically checked that the effect of the stochastic mechanism leads to a decrease in 2\% on solar neutrino survival probability, which would only marginally change the data analysis involving these neutrinos. For these reasons, we will focus our analysis just in experiments of very short baselines.

\section{The $\chi^2$ analysis}

We present the comparison of our proposal, the SNMM mechanism, with the sterile neutrino mechanism and the usual three neutrino scenario for the following data:  all set of old reactors (21 data point)~\cite{PhysRevD.83.073006},  SAGE and GALLEX calibration experiments (4 data point)~\cite{Anselmann1995440,Hampel1998114,PhysRevC.80.015807},  anti neutrino channel in LSND (8 data point)~\cite{PhysRevD.64.112007}, neutrino and anti neutrino channel in MiniBooNE (22 data point)~\cite{Aguilar-Arevalo:2013pmq} and  neutrino and anti neutrino channels in NuTeV (34 data point)~\cite{PhysRevLett.89.011804}, using a $\chi^2$ defined as:
\begin{equation}
\chi^2 = \sum_{i=1}^{ 89} {(R^t_i - R^e_i)}{W^{-1}_{ij}}{(R^t_j - R^e_j)},
\label{chi}
\end{equation}
where the $R^t$ are the theoretical predictions, $R^e$ are the experimental measurements, $W$ is the correlation function, and the sum is performed using 89 data points in total. 
For the  old-reactors data and for the GALLEX/SAGE data  we use off-diagonal elements of the correlation matrix from Ref.~\cite{PhysRevC.83.054615}.

\subsection{Comparing Sterile and SNMM}

We show our $\chi^2$ analysis results for all  data and three different scenarios: the 3+1 sterile model, the Stochastic Neutrino Mixing Mechanism and the three-neutrino case in Table \ref{res2}.  We present the best fit values for $\chi^2$ for each experiment and also for the full data set. 

For the three neutrino case, one cannot have short baseline oscillation using the standard values of mixing angles and mass differences for any of the experiments that we analyzed.  For the experiments that show no oscillation, such as NuTeV, there is a good agreement between data and theoretical predictions. But for other experiments, such as the reactors and the Mini-BooNE experiment,  there is a disagreement.  The $\chi^2$ has no free parameters and gives a very bad description of the data with the value of 148.14 with a very bad goodness-of-fit of  $8.4\times 10^{-5}$. 

\begin{table}
\centering
\vspace{.5cm}
\begin{tabular}{|c|c|c|c|c|c|}
    \hline
   &   $3\nu$   & 3+1 &  SNMM &  data points \\ \hline\hline
 Reactors &  34.41  & 22.58 &  30.16  & 21 \\ \hline
 SAGE /GALLEX & 8.09 & 5.26 &  3.27 &   4 \\ \hline
 LSND $\bar{\nu}$ & 16.48  &  3.77 &  3.89  & 8 \\ \hline
 MiniBooNE $\bar{\nu}$  & 18.69  &   6.98 &  15.54  & 11 \\ \hline
 MiniBooNE $\nu$& 28.56 & 11.76 & 20.81    & 11 \\ \hline
 NuTeV $\bar{\nu}$  & 25.32  & 25.32    &  25.84 & 17 \\ \hline
 NuTeV $\nu$  & 16.59 & 16.58 &  10.18   & 17 \\ \hline\hline
 Total & 148.14 & 92.24 &  109.72  & 89 \\ \hline
 d.o.f.	& 89 & 86 &  86  & - \\ \hline
 G.O.F. &  $8.4\times 10^{-5}$  & 0.303 &  0.043 & -  \\ \hline
     \end{tabular}
\caption{$\chi^2$ best fit values for each experiment for the combined sets for the 3+1  sterile model, for the stochastic mechanism and for the usual $3\nu$ case. We also show the degrees of freedom (d.o.f) and the goodness of fit (G.O.F.) for the combined data.}
\label{res2}
\end{table}

The  sterile neutrino parameters that produce the best fit to data for the 3+1 model are $\Delta m^2_{41}~=~0.42$~eV$^2$,   ${\rm U}_{\mu 4}~=~0.29$ and  ${\rm U}_{e 4}~=~0.14$. Our values are in reasonable agreement with the values of other 3+1 model analyses~\cite{Abazajian:2012ys}.  The goodness-of-fit value shows that the sterile neutrino case  is a reasonable explanation for the combined data with the best fit value providing $\chi^2_{\rm b.f.}=92.24$ for 86 d.o.f,  with a goodness-of-fit value of 0.303.

 \begin{figure}
		\includegraphics[scale=.34]{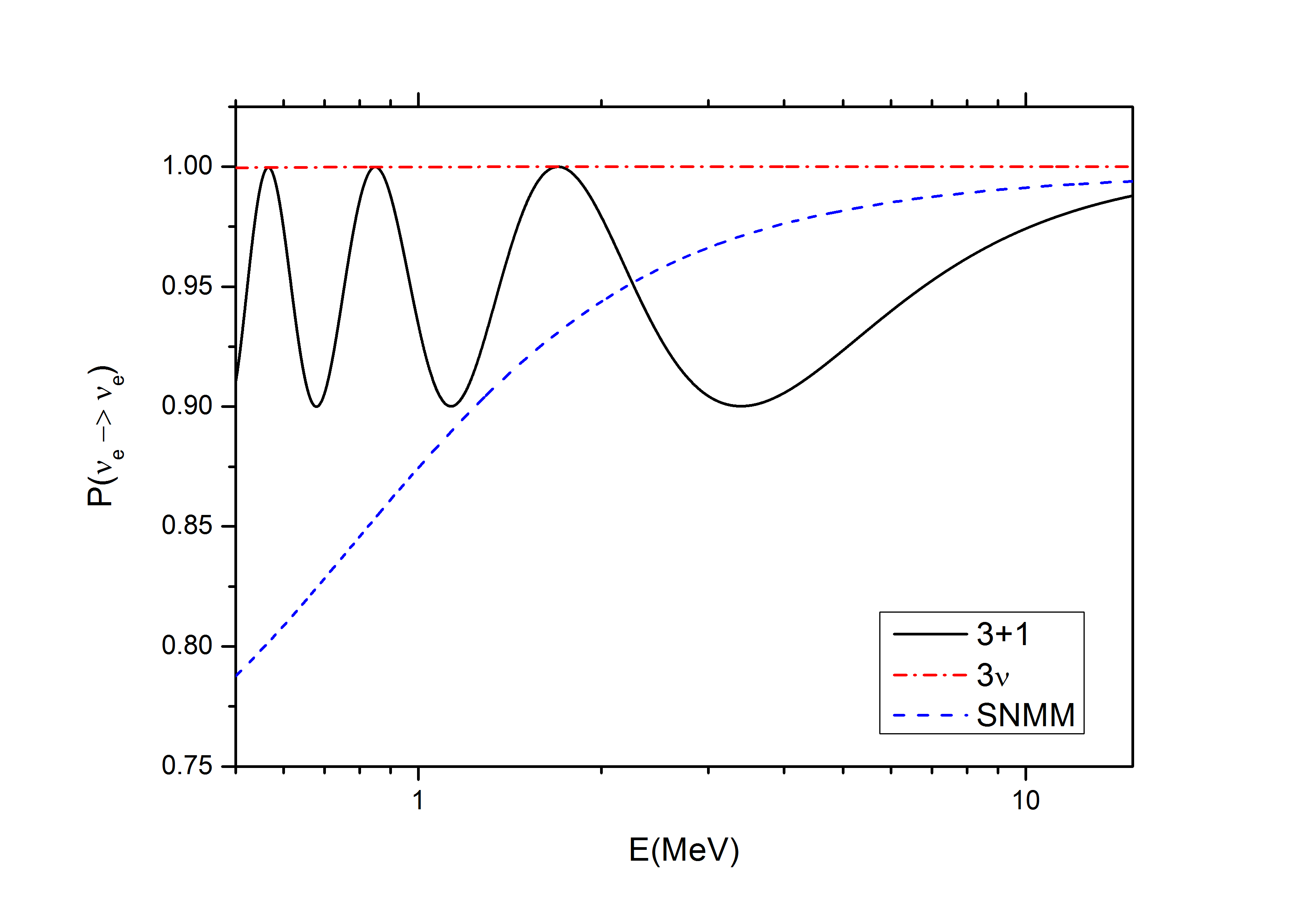}
		\caption{The black line represents the behavior of the 3+1 sterile neutrino model using the best fit values obtained in the global fit ($\Delta m^2_{41}~=~0.42$~eV$^2$, ${\rm U}_{\mu 4}~=~0.29$ and  ${\rm U}_{e 4}~=~0.14$), the red line represents the behavior of the standard 3$\nu$ scheme using the best fit values of \cite{PhysRevD.86.073012} and the blue line represents the behavior of the SNMM using the best fit values obtained in the global fit ($\alpha = A + B/E^n$, $A_{\rm b.f.}=0.026, B_{\rm b.f.}=0.26$ and $n_{\rm  b.f.}=0.80$). All of the results are obtained using a fixed distance of 10 m.
$P_{3+1}$ depends on $L/E$, and is straightforward to re-escalate the figure for different values of $L$, while the $P_{SNMM}$ has the same dependence with $L$ as $P_{3\nu}$, which is negligible for very-short baselines.}
		\label{plovere}
	\end{figure} 
For the Stochastic Neutrino Mixing Mechanism mechanism, the free parameters are the energy dependence of the gaussian width   $\alpha = A + B/E^n$.  We have found for the global analysis that  $A_{\rm b.f.}=0.026, B_{\rm b.f.}=0.26$ and $n_{\rm  b.f.}=0.80$ for neutrino energies given in MeV.
The SNMM fit is better than the $3\nu$ case, but it is not better than the sterile case. The main difference in $\chi^2$ between these two scenarios comes from the reactor analysis, mainly because the flat pattern reproduced by the SNMM does not fit so well the data as the rippled pattern of the sterile neutrino model. The same behavior also provides better results to the sterile neutrino in the region of MiniBooNE data. In order to clarify this we show in Fig~(\ref{plovere}) the behavior of the $3\nu$, the sterile and SNMM cases for the $P_{\nu_e \rightarrow \nu_e}$ as a function of the energy for length of 10m, which is a characteristic length for reactors experiments. In this figure the standard 3 $\nu$ model predicts none oscillation while the SNMM provide a conversion effect and the sterile neutrino have an oscillation effect. For the best fit values of sterile case and SNMM we have a different conversion pattern, the sterile case have a short oscillation length and the SNMM have a conversion effect due the averaging effect in Eq.~(\ref{snmm3fam1IIf}).





\section{Possible tests}

One of the possible tests to confirm or exclude the SNMM in comparison to the sterile neutrino models consists in allocating a source of $^8$Li inside a detector sensitive to charged and neutral current signals, such as SNO experiment~\cite{PhysRevLett.87.071301}, for example. The $^8$Li source
produces $\bar{\nu}_e$ neutrinos with the average energy of 6.4 MeV and we will assume a detector  located at most at 16m from the source.
 Using the SNO setup that consists in a tank filled with heavy water, reactions like $\bar{\nu}_e + D \rightarrow n + n + e^+$ and $\bar{\nu} + D \rightarrow \bar{\nu}' + n + p$, will happen.   For these configuration, with $L/E\sim 16/6.4 \sim 2.5 {\rm m/MeV}$,  the standard oscillation scenario predicts no-oscillation and any oscillation seen is due to new  physics. 

The original motivation for this experiment is to observe the oscillation pattern of the sterile neutrino model for the typical parameters found in previous section.  We suggest to extend the range of possible models to be tested in this experiment,  including the Stochastic Neutrino Mixing Mechanism, by a measurement of the  charged/neutral currents of $\bar{\nu}_e$. 

For this we computed the expected rate for the standard $3\nu$ scenario both for the charged current, $N^{\rm CC}$ as well for the neutral current $N^{\rm NC}$ as
\begin{equation}
N_{3\nu}^{\rm CC/NC}=\int{\frac{n_T T S_{^8{\rm Li}}}{L^2}
\sigma_{\rm CC/NC}dEdL},
\end{equation}
where the cross section ($\sigma_{\rm CC/NC}$) is obtained in~\cite{PhysRevD.60.053003} and  the spectrum ($S_{^8{\rm Li}}$) is provided by~\cite{PhysRevLett.109.141802}. Here $n_T$ is the number of targets and T is the lifetime.

For the sterile mass model, we expect to see oscillations, and then the number of events for charged and neutral currents  is modified as 

\begin{eqnarray}
\left(
\begin{array}{c}
N_{3+1}^{\rm CC} \\ 
N_{3+1}^{\rm NC}
\end{array}
\right)
=\int{\frac{n_T  T S_{^8{\rm Li}}}{L^2}\left(
\begin{array}{c}
\sigma_{\rm CC} \\ 
\sigma_{\rm NC}
\end{array}
\right)
\left(
\begin{array}{c}
 P_{\nu_e \rightarrow \nu_e}^{3+1} \\ 
 \sum_x P_{\nu_e \rightarrow \nu_x}^{3+1}
\end{array}
\right)
dEdL}.
\nonumber
\end{eqnarray}
 Here $\sum_x P_{\nu_e \rightarrow \nu_x}^{3+1}\neq 1$ due the presence of sterile neutrino.

For the Stochastic Neutrino Mixing Mechanism we have similar expressions,
\begin{eqnarray}
\left(
\begin{array}{c}
N_{SNMM}^{\rm CC} \\ 
N_{SNMM}^{\rm NC}
\end{array}
\right)
=\int{\frac{n_t T S_{^8{\rm Li}}}{L^2}\left(
\begin{array}{c}
\sigma_{\rm CC} \\ 
\sigma_{\rm NC} 
\end{array}
\right)
\left(
\begin{array}{c}
 P_{\nu_e \rightarrow \nu_e}^{\rm SNMM} \\ 
1
\end{array}
\right)
dEdL},
\nonumber
\end{eqnarray}
where in SNMM we have $ \sum_x P_{\nu_e \rightarrow \nu_x}^{\rm SNMM}=1$.
In the neutrino sterile hypothesis a detection rate decrease is expected in both neutral current and charged current channels due to the oscillation of electronic anti-neutrino in sterile anti-neutrinos. But in the SNMM hypothesis, while the charged current detection rate decrease due to the SNMM zero distance effect, the total number of active neutrinos remains constant, and then  the detection rate through neutral current remains the same as in the standard $3\nu$ oscillation model. 

To characterize these differences between the sterile model, the Stochastic Neutrino Mixing Mechanism and the standard $3\nu$ model we propose three observables:  The ratio of observed charged current events,  $N_{\rm obs}^{\rm CC}$  and the neutral current events, $N_{\rm obs}^{\rm NC}$ to the standard $3\nu$ case and the double ratio of NC over CC ratios  are defined as, respectively,  
\begin{equation}
\phi_1 = \dfrac{N_{\rm obs}^{\rm CC}}{N_{3\nu}^{\rm CC}}, \quad \phi_2 = \dfrac{N_{\rm obs}^{\rm NC}}{N_{3\nu}^{\rm NC}}, \quad \phi_4 = \dfrac{\phi_2}{\phi_1}.
\label{phi1}
\end{equation}
We will assume an  experiment running for 5 years with the detector of the size of SNO detector, 1 kton.  We can do two types of analysis, the rate and the shape of measured data for this configuration.

\begin{figure}
		\includegraphics[scale=.34]{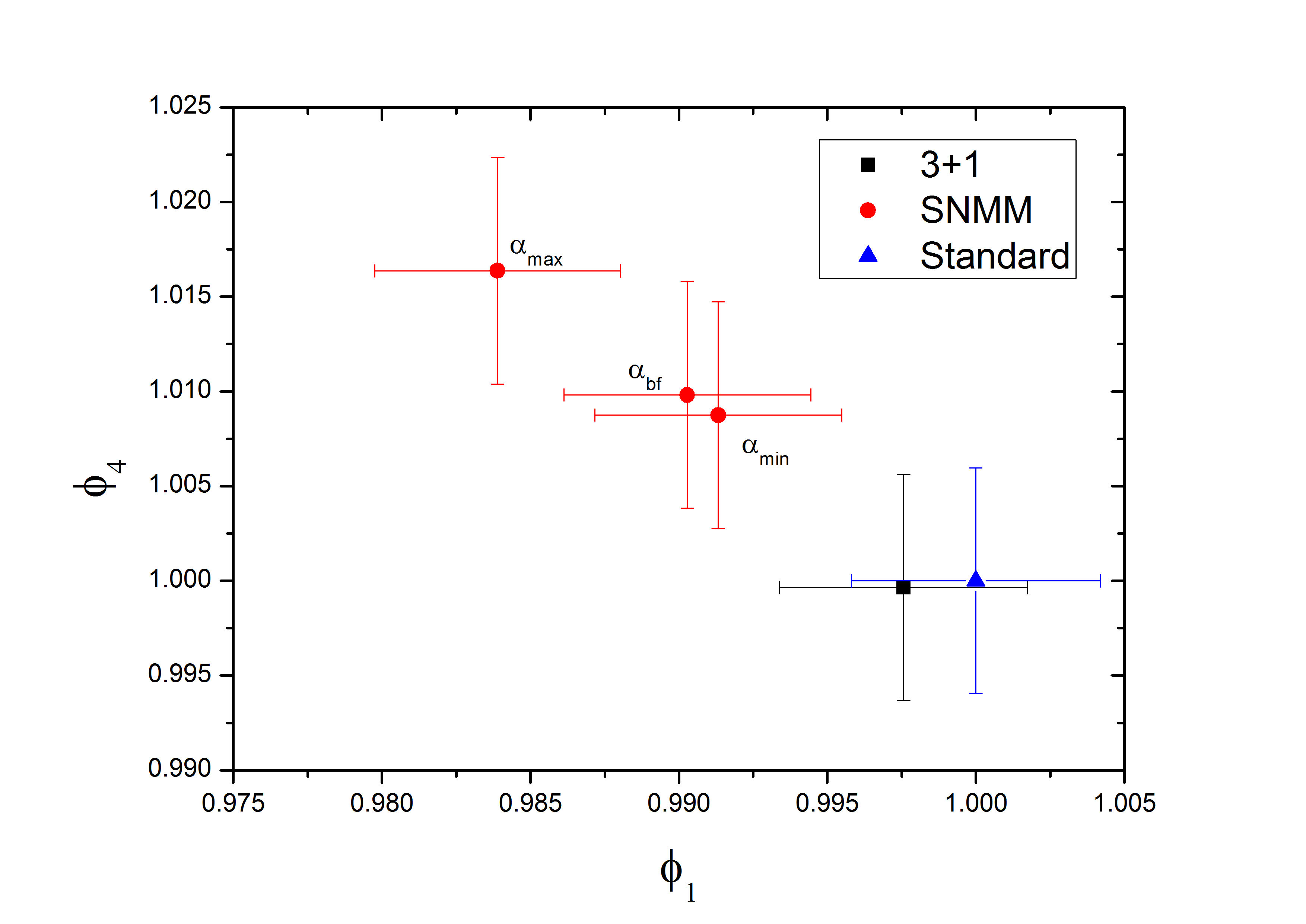}
		\caption{Ratios $\phi_1$ and  $\phi_4$ defined in Eq.~(\ref{phi1}).  To plot the 3+1 curves was used the best fit values of our sterile analysis. For the Stochastic Neutrino Mixing Mechanism we use the  $1\sigma$ values of A,B and n parameters of our parametrization of Gaussian width $\alpha = A + B/E^n$. The points indicated by $\alpha_{\rm max},\alpha_{\rm b.f.}$ and $\alpha_{\rm min}$  indicate the maximum, best fit values and minimum values for  Stochastic Neutrino Mixing Mechanism. }
	\label{fig:NC-CC}
\end{figure}
We show in Fig. \ref{fig:NC-CC} the predicted values for $\phi_1$ and $\phi_4$ for the best fit values of 3+1 model and for the Stochastic Neutrino Mixing Mechanism.  We can notice that from the rate only analysis we cannot discriminate between the standard $3\nu$ scenario and the best fit of 3+1 sterile model since the predictions overlap within $1\sigma$.  For the Stochastic Neutrino Mixing Mechanism, we have the $1\sigma$ values of free parameters
$\alpha = A + B/E^n$  of this mechanism to be $A_{\rm min}=0.02, A_{\rm max}=0.028$, $B_{\rm min}=0.26$, $B_{\rm max}=0.36$ and $n_{\rm min}=0.78$ and $n_{\rm max}=0.79$.  One can see that we can discriminate at  $2\sigma$ the standard $3\nu$ case from  the Stochastic Neutrino Mixing Mechanism.  Even more important, one can see from Fig.~(\ref{fig:NC-CC}) that for some values of the $\alpha$ parameter close to $\alpha_{\rm max}$, it is possible to distinguish SNMM from the 3+1 sterile model at $2\sigma$ in 5kton-yr of running.

This shape analysis was already performed by~\cite{PhysRevLett.109.141802} in a KamLand-like detector which has almost the same fiducial mass than SNO. Ref.~\cite{PhysRevLett.109.141802} concluded that in a five-year running experiment it is possible to distinguish between the 3+1 and the standard oscillation model in a KamLand detector size. Since the Stochastic Neutrino Mixing Mechanism will provide no spectrum shape changes, this test will not add any information.
A combined analysis (shape + rate) using CC and NC in the SNO experiment can definitively distinguish between the sterile neutrino model, the standard oscillation model or the SNMM.  As a disclaimer we use only statistical errors and the existence of systematic errors can change the conclusions. The rate $\phi_4$ is less sensitive to systematic errors and give a better perspective to discriminate between the  sterile neutrino and the stochastic models. \\

\section{Conclusions}

The Stochastic Neutrino Mixing Mechanism supplies an alternative explanation for the short baseline experiments that can explain together the positive oscillation signal of reactors, SAGE/GALLEX experiments, MiniBooNE and LSND experiments and the negative results of NuTeV. The agreement with data improve compared with standard $3\nu$ neutrino scenario, but not with the same quality of 3+1 model.

To distinguish the solution offered by the SNMM from that one originated from the sterile neutrino hypothesis we can use an artificial radioactive source and monitor the rate and shape of the spectrum of produced neutrinos. We propose the following variables, the CC ratio, $\phi_1 = \dfrac{N_{\rm obs}^{\rm CC}}{N_{3\nu}^{\rm CC}}$, the NC ratio $\phi_2 = \dfrac{N_{\rm obs}^{\rm NC}}{N_{3\nu}^{\rm NC}}$, and the NC/CC double ratio $\phi_4 = \dfrac{\phi_2}{\phi_1}$ to discriminate between the two scenarios.

In an experiment with production of electronic anti-neutrinos, if a decrease in the NC signal is observed, the SNMM will be excluded, otherwise, an evidence in favor of the SNMM will be found. A combined rate and shape analysis using an SNO-like detector can point out in favor of the SNMM or the Sterile Neutrino hypothesis in a 5kton-yr running experiment.

\begin{acknowledgments}
The authors would like to thank FAPESP, CNPq and CAPES for several financial supports. O.L.G.P. thanks the support of funding grant  2012/16389-1, S\~ao Paulo Research Foundation (FAPESP). E. Z. thanks the support of funding grants 2013/02518-7 and 2014/23980-3, S\~ao Paulo Research Foundation (FAPESP).
\end{acknowledgments}

\bibliography{references} 
\end{document}